\documentstyle[emulateapj,onecolfloat,psfig]{article}   

\begin{document}
\twocolumn[      

\title{Distances and Cosmology From Galaxy Cluster CMB Data}
\author{Asantha Cooray}
\affil{Theoretical Astrophysics, Mail Code 130-33, California Institute of Technology, Pasadena, CA 91125\\
E-mail:asante@tapir.caltech.edu}


\begin{abstract}
The measurement of angular diameter distance to galaxy clusters, through
combined Sunyaev-Zel'dovich (SZ) effect data with X-ray emission
observations, is now a well-known probe of cosmology. Using a combination
of SZ data and a map of the lensed CMB anisotropies by  the galaxy cluster
potential, we propose an alternative geometric technique to measure
distance information primarily through cluster related multi-frequency
CMB measurements.  We discuss necessary requirements to implement this
measurement, potential errors including systematic biases, and the extent
to which cosmological parameters can be extracted. While individual
cluster distances are not likely to be precise, with upcoming subarcminute
resolution wide-area CMB observations, useful information on
certain cosmological parameters,  such as the equation of
state of dark energy, can be obtained from a large sample of galaxy
clusters.
\end{abstract}
\keywords{cosmology:theory --- large scale structure of the universe --- gravitational lensing ---cosmic microwave background}
]

\section{Introduction}

The measurement of distances to standard, or standardizable, candles and/or rulers at a known redshift 
provides a well-known probe of physical cosmology 
(see, \cite{Pee93} for a review). The well utilized techniques, so far, include luminosity distance to Type Ia 
supernovae (e.g., \cite{Rieetal98,Peretal99}), angular diameter distance to horizon size at the last scattering (e.g., \cite{Kametal94}), time delay between gravitationally-lensed images (e.g., \cite{Ref64}),
and information from clusters related to the Sunyave-Zel'dovich (SZ; \cite{SunZel80}) effect combined with X-ray emission observations
(e.g., \cite{Reeetal02}), among others.
Here, we propose an additional geometric test related to galaxy cluster CMB data involving maps of the SZ effect and
lensed CMB anisotropies.

The proposed technique makes use of the fact that the background source, primordial CMB anisotropy, lensed by a foreground cluster
has well-known properties, including the distance to the last scattering surface where the source is located. 
This is significant given that most lensing studies based on source
ellipticities or flux measurements, such as mass reconstruction of galaxy clusters via weak lensing shear, is affected by the unknown
redshift distribution of the  lensed background galaxy population.
 Given a map of the lensed CMB anisotropy distribution, the technique allows one to extract cosmological information
to the accuracy that the foreground mass distribution is known a priori. While this has, traditionally, come from 
lensing studies at optical wavelengths, we suggest, in principle, that the cluster SZ map can be used as a estimator of the mass 
distribution under certain assumptions on how the gas mass, which contributes to the SZ effect, is related to the total matter 
responsible for lensing.

The use of SZ and lensed CMB data 
is motivated by the fact that in the near future, with instruments such as the South Pole Telescope (SPT), and in the long term,
with missions such as the CMBpol, one expects large catalogs of arcminute scale CMB data related to galaxy clusters. 
In the case of multi-frequency CMB data, the cluster SZ effect can be separated based on its unique frequency spectrum relative to the
thermal CMB. When the dominant SZ effect is frequency-separated, the anisotropy towards an individual cluster is likely to contain 
two sources of importance: the SZ kinetic effect resulting from the peculiar velocity of the galaxy cluster along the line of 
sight and the lensing effect (\cite{SelZal00}).  The two effects have distinct spatial distributions while the
frequency-cleaned cluster SZ map provides a filter to remove the kinematic SZ contribution. This is due to the fact the two
SZ effects are expected have the same spatial distribution as they both trace the electron density field. When removed, the kinematic 
SZ data provide a measure of large scale bulk flows and a separate measure of cosmology (\cite{PeeKno02}). 
Here, we address cosmological uses of the combined 
SZ and lensed CMB maps of a cluster and extend that to a sample of clusters.
While an individual cluster distance  measured from the SZ/lensing method is unlikely to be statistically significant, 
for a large sample of galaxy cluster ($\sim 10^4$), we show that one can obtain detailed information related to cosmology.

The discussion is organized as follows. In the next Section, we discuss important aspects related to the proposed test
and practical considerations related to how well the technique can be implemented in future data.
In Section~3, we discuss cosmological measurements and conclude with a summary. 

\section{Cosmological Information via SZ and Lensing}

\begin{figure*}[t]
\centerline{\psfig{file=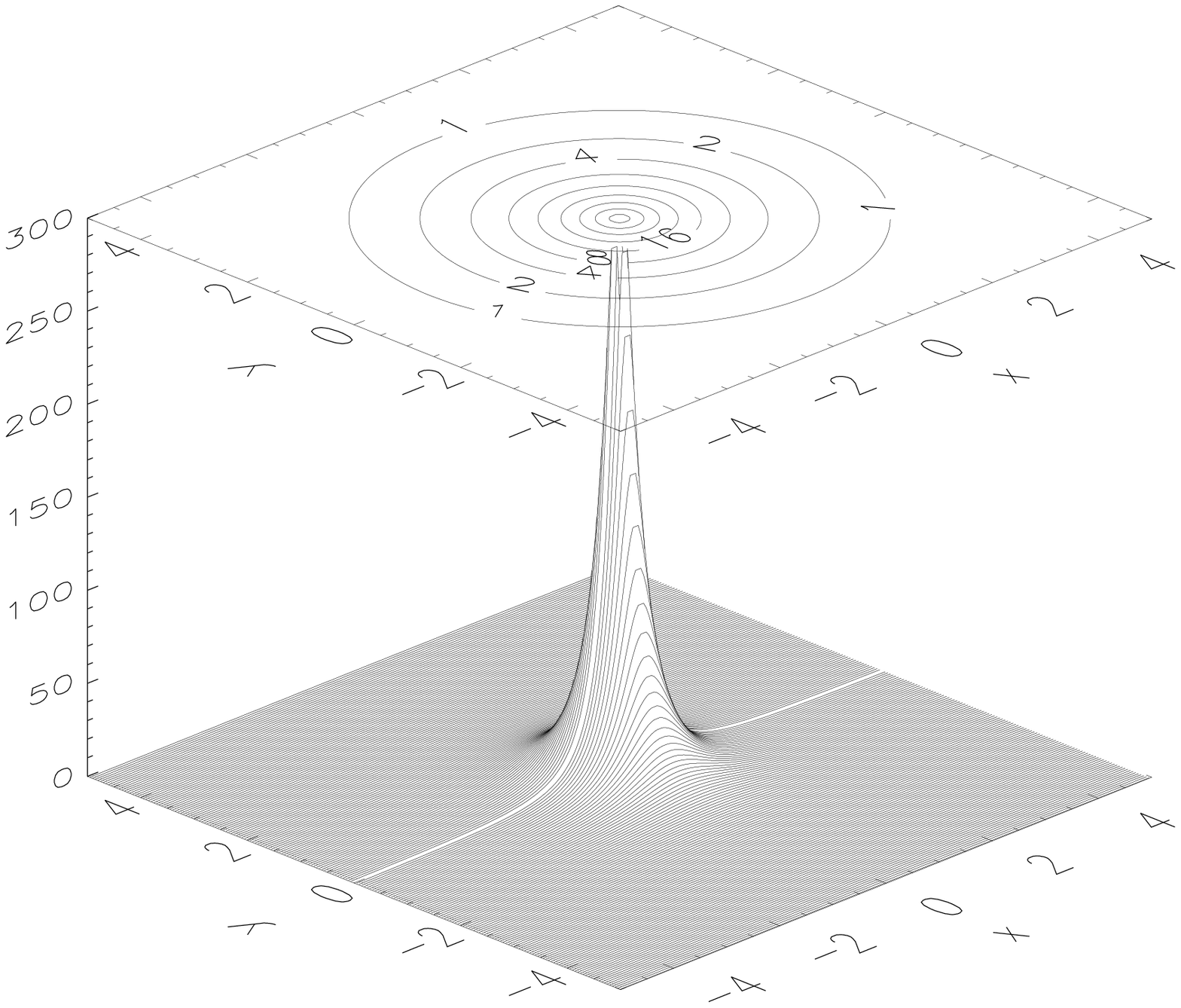,width=2.8in,angle=90}\psfig{file=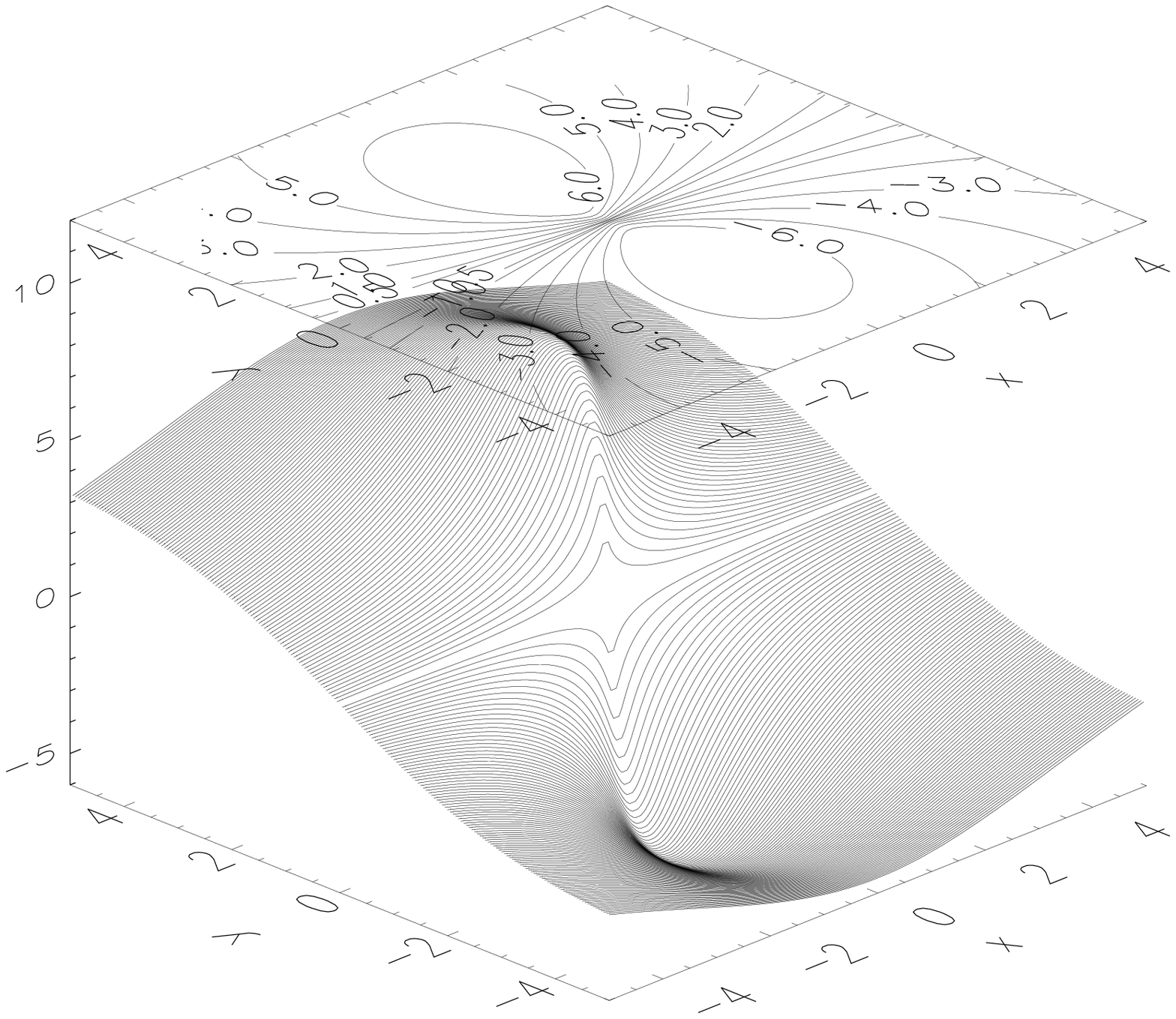,width=2.8in,angle=90}\psfig{file=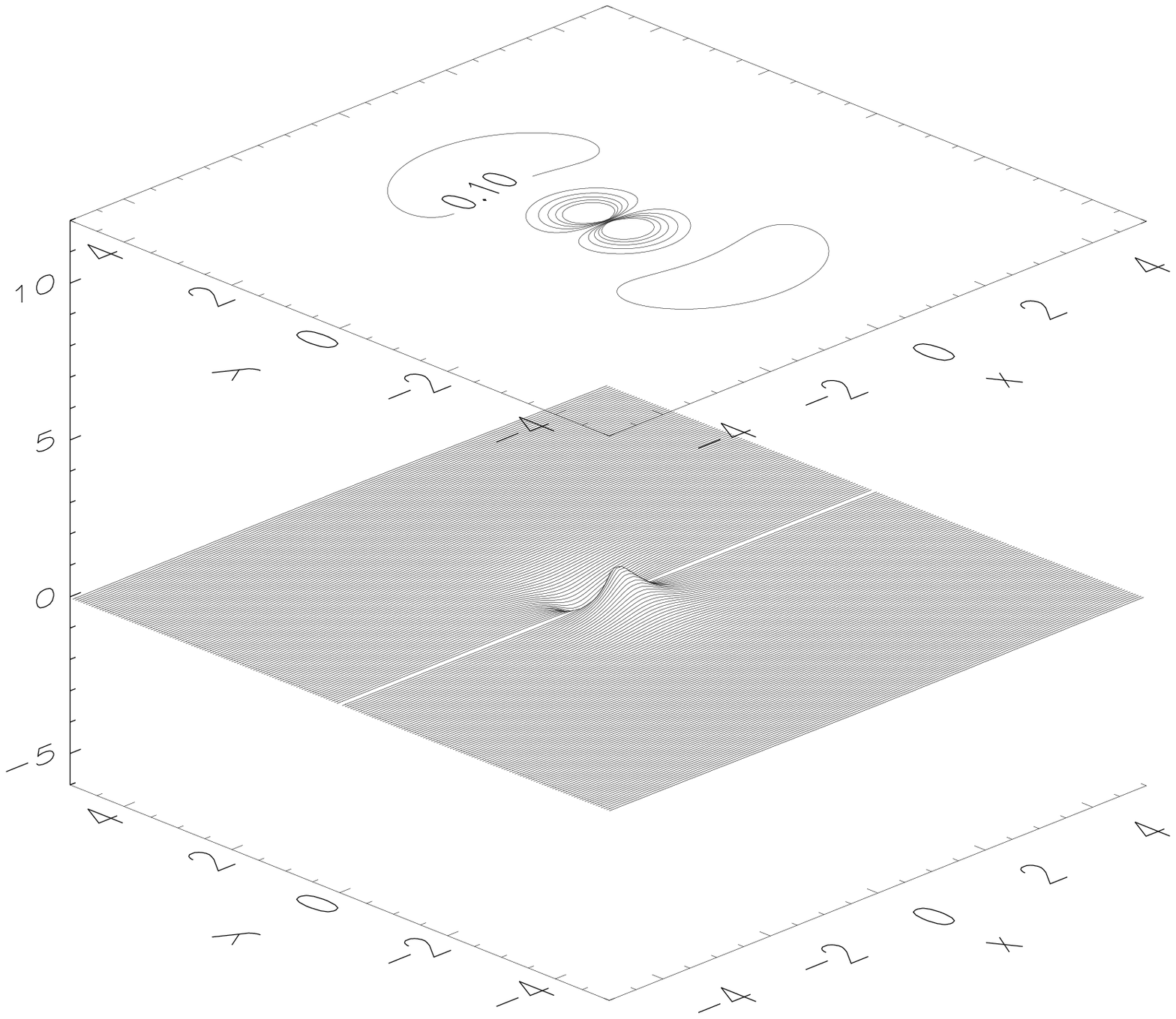,width=2.8in,angle=90}}
\caption{The SZ (a) and lensed CMB (b) maps of a 5 $\times$ 10$^{14}$ M$_{\sun}$ galaxy cluster at a 
redshift of 0.5. The coordinates in $x$ and $y$ axis are distances from the cluster center in terms of the scale radius while
the $z$-axis is temperature fluctuations in units of $\mu$K.
The SZ profile is calculated under the assumption that the gas distribution is in hydrostatic equilibrium with
the dark matter distribution. In (c), we show the difference between the  lensed CMB map (b) and the prediction based on an estimator of the
total mass that make use of the SZ profile in (a). The difference is minor when one compares at a large radius from the cluster center 
($\lesssim 0.1$ $\mu$K), while departures of order 0.5 $\mu$K are observed near the cluster center. This is due to the fact that
the cluster gas profile flattens near the inner regions of the cluster, when compared to the dark matter distribution.} 
\label{fig:profile}
\end{figure*}

The SZ effect measures the total integrated pressure along the line of sight through the cluster:
\begin{equation}
{\rm SZ}(\theta) = g(x) \frac{\sigma_T}{m_e c^2} \int d \ell \;  p_e(\theta d_L,\ell) \, \nonumber \\
\end{equation}
where $p_e(R,z)$ is the pressure profile, where $R=\theta d_L$ is the projected distance from the cluster center, at
a distance of $d_L$,  and $\ell$ is the line of sight distance. 
The SZ effect, as is now well-known, is frequency dependent through the spectrum, relative to the CMB blackbody spectrum, given by
$g(x)=x \coth(x/2)-4$,  where $x=h\nu/kT_{\rm CMB}$. This frequency dependence is important since it allows the SZ effect to 
be safely separated from both sources of interest, such as the kinetic SZ effect and lensed CMB anisotropies,
and confusions. The area integral of the frequency cleaned SZ map provides an estimator of the gas mass enclosed within the angular radius,
 weighted by the spatially averaged temperature distribution of electrons:
\begin{equation}
{\hat M}_{\rm gas}(\theta) = \frac{m_e m_p \mu_e c^2}{\sigma_T} 
\frac{d_L^2}{k_B {\tilde T}_e}  \int_0^\theta d^2\theta' {\rm SZ}(\theta') \, ,
\end{equation}
where $\mu_e$ is the electron fraction in the cluster and ${\tilde T}_e$ is the average electron temperature.  
The factor $d_L^2$, where $d_L$ is the cluster distance, comes from the conversion of physical distances to angular units.

Due to a variety of reasons, including the detection of inflationary gravitational wave signature in CMB data 
(\cite{Kesetal02,KnoSon02}), there is now a significant attempt to understand the gravitational lensing effect on CMB anisotropies.
Since surface brightness is conserved, the lensing effect on CMB is simply a modification to the photon distribution on the sky. 
In general, through a geometric factor related to the background source distance ($d_S$), 
in our case CMB at  the last scattering surface, and the distance to the foreground lens ($d_L$), which in this case is a
 galaxy cluster, the amplitude of the gravitational lensing effect depends on the background cosmology.
For the lensed-CMB case, angular deflections of the CMB photons lead to a modification to the large scale CMB gradient such that
\begin{eqnarray}
&& \Delta T_{\rm lens}(\theta) = \nabla (\Delta T_{\rm prim}) \cos \phi \frac{4G}{c^2} \frac{d_{LS}}{d_S d_L\theta} M_{\rm dm}(\theta) \, ,
\end{eqnarray}
where the effect is now sensitive to the total enclosed mass, which we take to be due to the dark matter distribution,
and $\cos \phi$ is the position angle with respect to the large scale CMB gradient, $\nabla (\Delta T_{\rm prim})$
(\cite{SelZal00,DodSta03}). Here, $d_{LS}$ is the distance between cluster and the background source.
Equation~(2) states that when $\Delta T_{\rm lens}(\theta)$ is measured, and $M_{\rm dm}(\theta)$ is either known or also
measured independently, information related to cosmology can be extracted from the geometric parameter  related to distances.
This is possible due to two reasons: $\nabla (\Delta T_{\rm prim}) \cos \phi$, 
is essentially known with a high signal-to-noise ratio from a map of the large scale CMB temperature fluctuations, 
such as from WMAP or Planck, while $d_S$ is also known from first principles and measurable from the CMB anisotropy angular power spectrum.

While $M_{\rm dm}(\theta)$ can be obtained from a variety of observational methods, such as lensing studies at optical wavelengths
based on background galaxy ellipticities and magnifications or estimates based on the X-ray emission,  we consider CMB 
related estimates of mass. This is motivated by the wealth of CMB related data expected in the near future 
with arcminute-scale CMB experiments, both from ground (SPT) and space (CMBpol).
The SZ effect allows a reliable estimate of the dark matter distribution under certain assumptions: the simplest
is to assume that the gas-to-total mass ratio traces that of the global baryon fractions such that
$M_{\rm dm}(\theta) = \eta^{-1} M_{\rm gas}(\theta)$ where $\eta=\Omega_b/\Omega_m$. While we do not consider in detail, 
one can potentially improve this assumption further based on numerical simulations. When SZ maps are used as
an estimator of the enclosed dark matter mass, the cosmological information comes from the combined parameter $f(z) \equiv 
d_{LS} d_l/d_S \; \Omega_m/\Omega_b$. 

When SZ data are used, note that we also need information related to
${\tilde T}_e$, average electron temperature, for individual clusters. While this can potentially come from 
X-ray observations, say, from wide field catalogs such as those expected from the proposed Dark Universe Explorer Telescope (DUET), 
the SZ spectrum can be used to extract an average value of the gas temperature based on the
relativistic corrections to the SZ effect (\cite{Poietal98}). This determination is best, at the level of 10\% to 15\% with Planck,
for massive clusters where the electron temperature is high ($>$ 5 keV), and these are the same clusters where the
lensing effect is expected to dominate. For a large sample of clusters, another approach would be to
potentially calibrate some relation between ${\tilde T}_e$ and the SZ effect and use such relations (\cite{Coo99,McC03}), 
combined with lensed CMB observations, to constrain cosmological parameters.

In Fig.~1, we illustrate the expected profile of the SZ effect (a) and the lensed CMB (b) towards a galaxy cluster. To calculate these maps,
we make use of the NFW profile of Navarro et al. (1996) to describe the dark matter distribution. We assume that the gas 
distribution is in hydrostatic equilibrium with the cluster potential. This allows us to calculate the pressure profile related 
to the SZ effect following Komatsu \& Seljak (2001). In Fig.~1(c), we show the difference between the lensed CMB map and 
the one predicted based on the integrated SZ profile. While the difference is minor at outer regions of the cluster, 
we find substantial differences near the inner part of the cluster. This is 
associated with the fact that gas profile does not trace dark matter distribution in the inner regions due to additional 
pressure support. When calculating the lensed CMB map, to compare with Fig.~1(b), note that we have 
scaled the SZ profile by the cluster-averaged value of the gas temperature
instead of using the exact temperature profile. While improvements to Fig.~1(c) is clearly expected when the temperature profile  
is included, we suggest that this is not necessary, at least, in terms of models we have considered. 
When one compares at several scale radii from the cluster center, 
we also found the difference with true lens profile and the predicted one to be below the expected noise levels of upcoming 
CMB observations. Note that the SZ/X-ray route to determine cluster distances require precise temperature information, including the profile, since
one squares the SZ map to be compared with X-ray data, thus increasing small differences, and for the reason that temperature
profile also enters the X-ray description. 
We expect further numerical  studies to address the extent to which biases can be introduced by
representing the temperature distribution with an average value in the SZ/lensing method.

\section{Cosmological Measurements}

Now to understand how well cosmological information can be extracted, we make use of the expected sample of galaxy clusters
detectable from a SPT-like experiment in its planned wide field survey of 4000 deg$^2$. 
These observations are expected to allow detection of clusters down
to a mass limit of $\sim 3 \times 10^{14}$ M$_{\sun}$, though we only make use of the cluster catalog with masses above
$5 \times 10^{14}$ M$_{\sun}$. Since noise levels and other important experimental aspects are still not well defined
for upcoming observations, in order to obtain some guidance, we assume a final pixel noise of 0.5 $\mu$K at one arcminute resolution
and assume multi-frequency observations such that the SZ effect is separated with an increase in the noise associated with
the SZ map by a factor of 2. This is roughly equivalent to an experiment with five channels equally separated 
from 217 GHz, which is the SZ null frequency. 

With SZ separated, to extract the lensed CMB map, we remove the kinetic SZ contribution by assuming that its contribution is 
simply related to the SZ profile with an overall amplitude given by the peculiar velocity. This removal is aided by the fact 
that the kinetic SZ and lensed CMB contributions have distinct spatial distributions. Following Sheth \& Diaferio (2001),
we assume a Gaussian distribution for cluster peculiar velocities with a zero mean, and a variance of 400 km sec$^{-1}$, 
consistent with $\Lambda$CDM  cosmology, and ignore virial motions within clusters since such motions do not contribute to 
the kinetic SZ effect due to cancellations. Any coherent motions, such as bulk rotations of cluster gas 
lead to an additional kinetic SZ related contribution (\cite{CooChe02}), though, we ignore such possibilities here since they are expected to be small, 
except in certain favored cases. For the lensing extraction, we only make use of the subsample of
clusters where the kinetic SZ signal is at the same level as the expected lensed CMB contribution and is expected to be
separated with no significant residual noise. Allowing for some accounting of  the
fact that the background temperature gradient also varies on the sky, such as clusters that lie on hot or cold-spots will not show
a significant lensing effect, we found roughly 10\% of the cluster catalog will be useful for the
proposed study. For a ground-based survey of
$\sim$ 10\% of the sky (with SPT), this results in  catalog of roughly 10$^3$ clusters useful for the proposed study. 

The lensed CMB contribution, for the most favorable clusters, 
is generally detected with signal-to-noise ratios, integrated over the whole cluster, at most at the level of ten.
Here, we have ignored contaminants such as radio point sources and dust since these are likely to be removed based on
frequency information. When the point source contribution is modeled with counts based on Toffolatti et al. (1999) models, 
and assuming point sources above 3-sigma noise level can be detected and removed, we found residual noise contributions at the 
level of 1 $\mu$K at 217 GHz, which is at the level of noise in the lensed CMB map. The main source of worry here is an extra 
population of sources that happen to reside in clusters, though, it is unlikely that such a population dominates at 
frequencies such as 217 GHz. 

Given a carefully selected subsample of clusters where one has reconstructed lensed CMB maps, and using SZ maps as estimates 
of the gas mass  distribution, we now calculate how well the geometric factor related to distances can be extracted.
For this purpose, we assume that the background CMB gradient is precisely known. This is a safe assumption given that experiments 
like WMAP and Planck produce high signal-to-noise maps of the large scale CMB temperature fluctuations. We assume for the 
subsample of clusters, the average  electron temperature of each cluster, either based on the SZ spectrum, or some statistical 
relation between the temperature and the SZ effect, is known to an accuracy of 10\%.
With improvements on CMB experimental side,  it is likely that this will be the main 
limitation for the proposed study. We illustrate our results in Fig.~2, where we show the function $f(z)$ and expected errors 
in its measurement. For the illustration purposes here, we bin  the cluster catalog assuming that their redshifts are known 
precisely.  For a given cluster, $f(z)$ is only known to an 
accuracy of, in general, 30\% and is usually not better than 15\% even in the most favored case. For a sample of 1000 clusters, 
binned in redshift out to a redshift of 2 in 10 bins, we find that $f(z)$ can be determined to an accuracy of order to 
$\sim$ 10\% in low and high redshift bins and at the level of 3\%  at a redshift of near unity  where cluster counts peak. 
This can be compared to the measurement errors on the Type Ia supernovae luminosity distances expected from the SNAP mission, which is at the level 
of 1\% in each redshift bin of width 0.1. While Fig.~2(a) shows the level expected for a survey of 4000 deg$^2$, for an all sky 
survey, as expected from the CMBpol mission, the useful catalog of clusters will increase by a factor $\sim$ 10, and the error 
on binned $f(z)$ reconstructed from such data is expected to improve due to the increase in the cluster sample.

In order to illustrate the extent to which cosmological measurements can be extracted with reconstructed $f(z)$, we make
use of the Fisher information matrix. We assume a flat-cosmological model where the dark energy contribution
has an equation of state $w$ and consider the measurement of two parameters related to $\Omega_m$ and $w$.
In Fig.~2(b), we show the expected errors and a comparison to the expected error ellipse on the same parameter plane 
from SNAP data. While a SPT-like ground-based survey provides limited cosmological information from the proposed test, increasing 
the capability to an all-sky survey, as expected from the CMBpol mission, one can extract cosmological information at the
same precision as the SNAP mission. 

\begin{figure}[t]
\centerline{\psfig{file=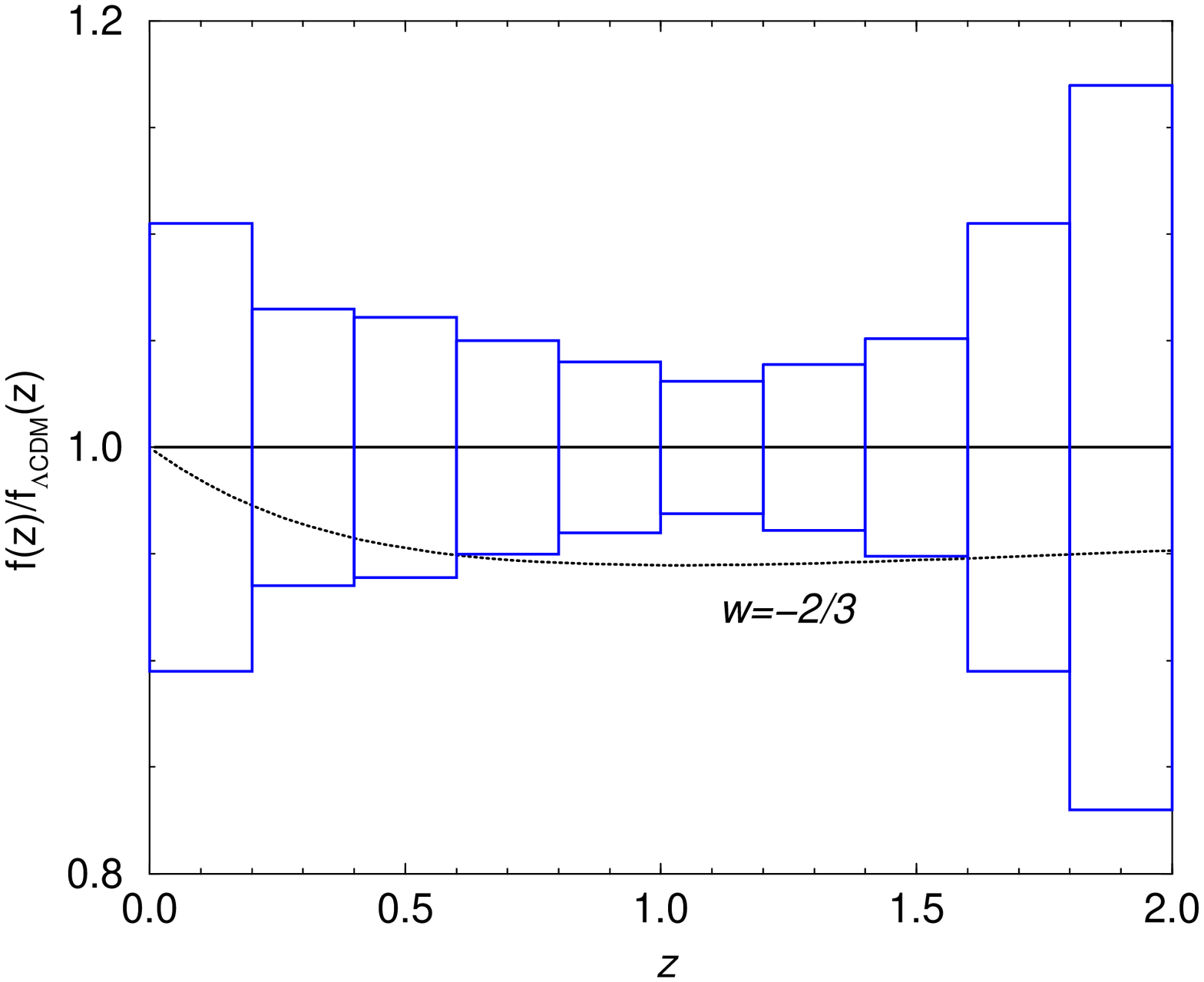,width=3.2in,angle=-90}}
\centerline{\psfig{file=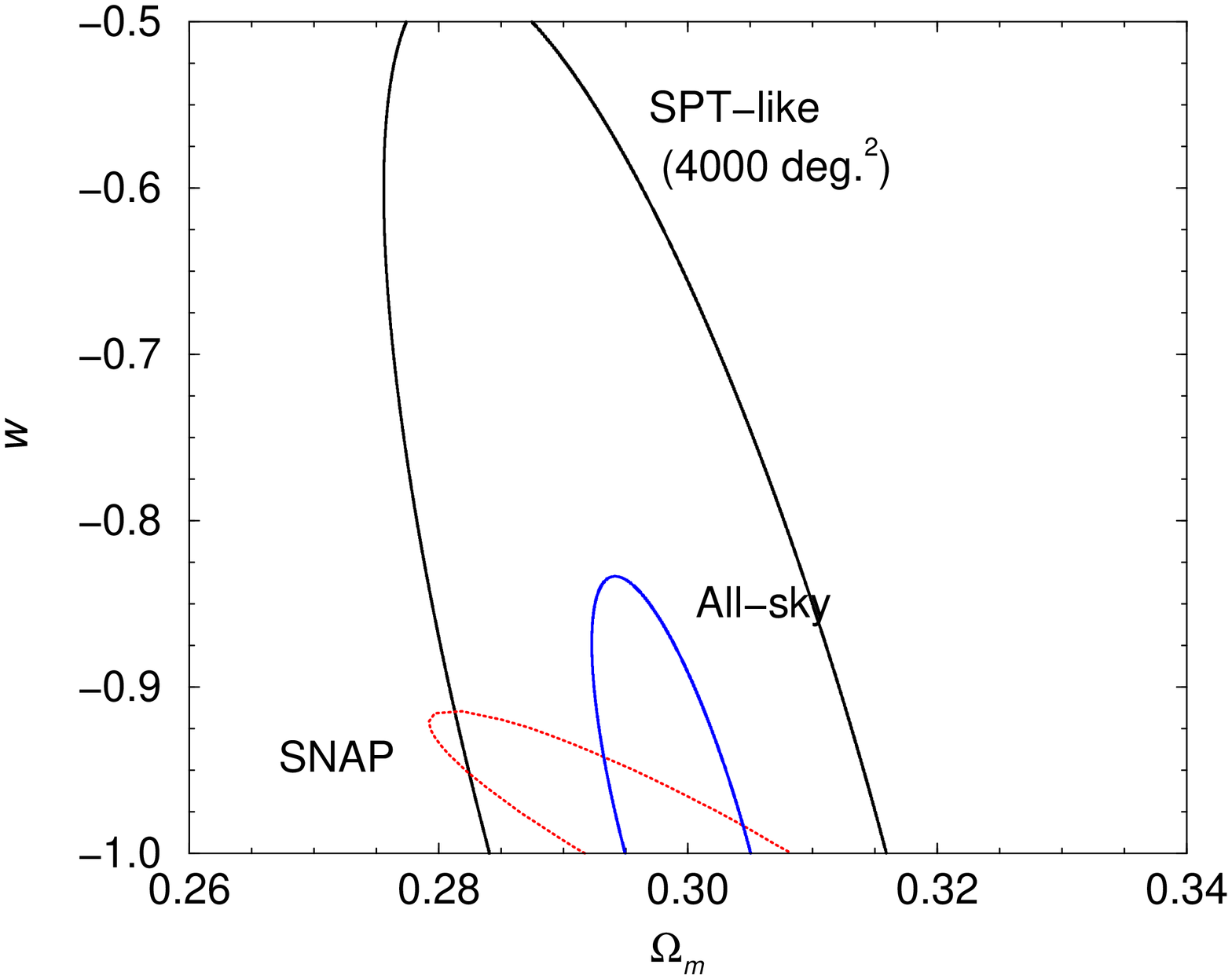,width=3.2in,angle=-90}}
\caption{(a) The expected errors on the function $f(z)$ from a subsample of 10$^3$ clusters useful for an extraction of the lensed CMB map. The dotted line shows the same quantity in a flat-cosmological model where the equation of state of the dark energy, $w$, is taken to be -2/3. Such a cosmology is distinguished by these measurements at the level of  few sigma. (b)
Cosmological parameter errors on $\Omega_m$ and $w$ from combined lensed CMB and SZ maps from a ground-based SPT-like
experiment (solid line) and an all-sky space-based experiment similar to CMBpol (dotted line). For comparison, we also
show the expected error ellipse from the SNAP mission. The error ellipse from SNAP and  the proposed test have different directions
suggesting the complimentary nature of the two tests.}
\label{fig:fz}
\end{figure}

To summarize, in the near-future, arcminute scale CMB surveys are expected to detected many tens of thousands
or more galaxy clusters based on the SZ effect. We suggest that a subsample of clusters ($\sim$ 10\%) will allow
useful extraction of information related to the CMB lensing effect and can be used with SZ data for an additional
cosmological test.  The proposed test makes use of the lensing geometric factor related to lens and source distances to extract distance 
information and associated cosmological parameters. While individual distance measurements are not significant, for a subsample 
of preselected clusters, such as based on the amplitude of $\nabla (\Delta T_{\rm prim})$ with respect to the cluster location,
we have shown that precision cosmology can be achieved. 

Unlike distances based on the SZ/X-ray route, the proposed cosmological test is independent of assumptions related to the cluster shape
including asphericity. This can be understood based on the fact that the effect relies on the line of sight projected quantities and 
not a line of sight projected quantity (SZ effect) combined with volume integrated emission (X-ray flux). 
The test, however, assumes some relation between gas mass and the dark matter mass of a cluster and the extent to which
this can be achieved should be studied with simulations. Even if not used for cosmology, we suggest conducting this study in
future data as a way to understand certain astrophysics related to galaxy clusters.

\smallskip
{\it Acknowledgments:} 
We thank S. Dodelson and G. Starkman for a preprint of their paper, which motivated the author to write this work, 
D. Huterer for SNAP information, and acknowledge partial support from the Sherman Fairchild Foundation and DOE.

\end{document}